\documentclass[aps,prl,reprint,twocolumn,superscriptaddress]{revtex4-2} 
\usepackage{amsfonts}
\usepackage{amssymb}
\usepackage{amsthm,amsmath}
\usepackage{mathrsfs}
\usepackage{indentfirst}
\usepackage{multirow}
\usepackage{graphicx}
\usepackage{amsmath}
\usepackage[amssymb]{SIunits}
\usepackage{amssymb}
\usepackage{color}
\usepackage{xcolor}
\usepackage{natbib}
\usepackage[T1]{fontenc}
\usepackage{bm}
\usepackage{booktabs}
\usepackage{multirow}
\usepackage{setspace}

\usepackage{ulem}
\usepackage[colorlinks=true, linkcolor=blue, citecolor=blue, urlcolor=blue, bookmarks=true]{hyperref}

\begin{document}

\title{How Meandering Flow Sculpts Ice}

\author{Yihong Du}
\affiliation{New Cornerstone Science Laboratory, Center for Combustion Energy, Key Laboratory for Thermal Science and Power Engineering of MoE, Department of Energy and Power Engineering, Tsinghua University, China}

\author{Jin Hu}
\affiliation{New Cornerstone Science Laboratory, Center for Combustion Energy, Key Laboratory for Thermal Science and Power Engineering of MoE, Department of Energy and Power Engineering, Tsinghua University, China}

\author{Feng Wang}
\thanks{Corresponding author: fengwang2023@tsinghua.edu.cn}
\affiliation{New Cornerstone Science Laboratory, Center for Combustion Energy, Key Laboratory for Thermal Science and Power Engineering of MoE, Department of Energy and Power Engineering, Tsinghua University, China}

\author{Lars ten Hacken}
\affiliation{Department of Applied Physics and Science Education, Eindhoven University of Technology, 5600 MB Eindhoven, Netherlands}

\author{Ziqi Wang}
\affiliation{Department of Applied Physics and Science Education, Eindhoven University of Technology, 5600 MB Eindhoven, Netherlands}

\author{Rui Yang}
\affiliation{Department of Engineering Mechanics, School of Aerospace Engineering, Tsinghua University, China}

\author{Enrico Calzavarini}
\affiliation{Unit\'e de M\'ecanique de Lille - J. Boussinesq (UML) ULR 7512, Universit\'e Lille, F-59000 Lille, France}

\author{Federico Toschi}
\affiliation{Department of Applied Physics and Science Education, Eindhoven University of Technology, 5600 MB Eindhoven, Netherlands}
\affiliation{CNR-IAC, Via dei Taurini 19, 00185 Rome, Italy}

\author{Chao Sun}
\thanks{Corresponding author: chaosun@tsinghua.edu.cn}
\affiliation{New Cornerstone Science Laboratory, Center for Combustion Energy, Key Laboratory for Thermal Science and Power Engineering of MoE, Department of Energy and Power Engineering, Tsinghua University, China}
\affiliation{Department of Engineering Mechanics, School of Aerospace Engineering, Tsinghua University, China}

\date{\today}

\begin{abstract} 

Ice-water phase change subject to a mean flow widely exists in the Cryosphere, with implications on glacier evolution, human transportation, flooding events, etc.
Understanding the relevant phase change dynamics requires identifying the roles of shear and rotation due to the mean flow.
Combining experiments, numerical simulations and theoretical modelling, we reveal the combined effects of shear and rotation on the coupled dynamics of water flow, heat transfer and phase change in an annular system.
Increasing shear and rotation drive a stronger secondary circulation which significantly enhances global heat transfer and redistributes the radial heat flux, and eventually lead to a thinner equilibrium ice layer and a larger ice thickness difference in the radial direction.
Further, we develop an integral thermal balance model that explains the mean ice thickness through estimating the additional heat transfer, and a local thermal balance model that explains the inner-outer ice thickness difference by relating it to the non-uniform convective heat flux. 
The local model shows qualitative agreement with field measurements when applied to an ice-covered meandering river.
These results contribute to understanding the common physical mechanism for ice-water phase change under shear and rotation, offering a theoretical basis for explaining and predicting ice evolution in meandering rivers among other relevant scenarios.

\end{abstract}

\maketitle

\section{I. Introduction}

\begin{figure*}[!htb]
\centering
\includegraphics[width=0.85\linewidth]{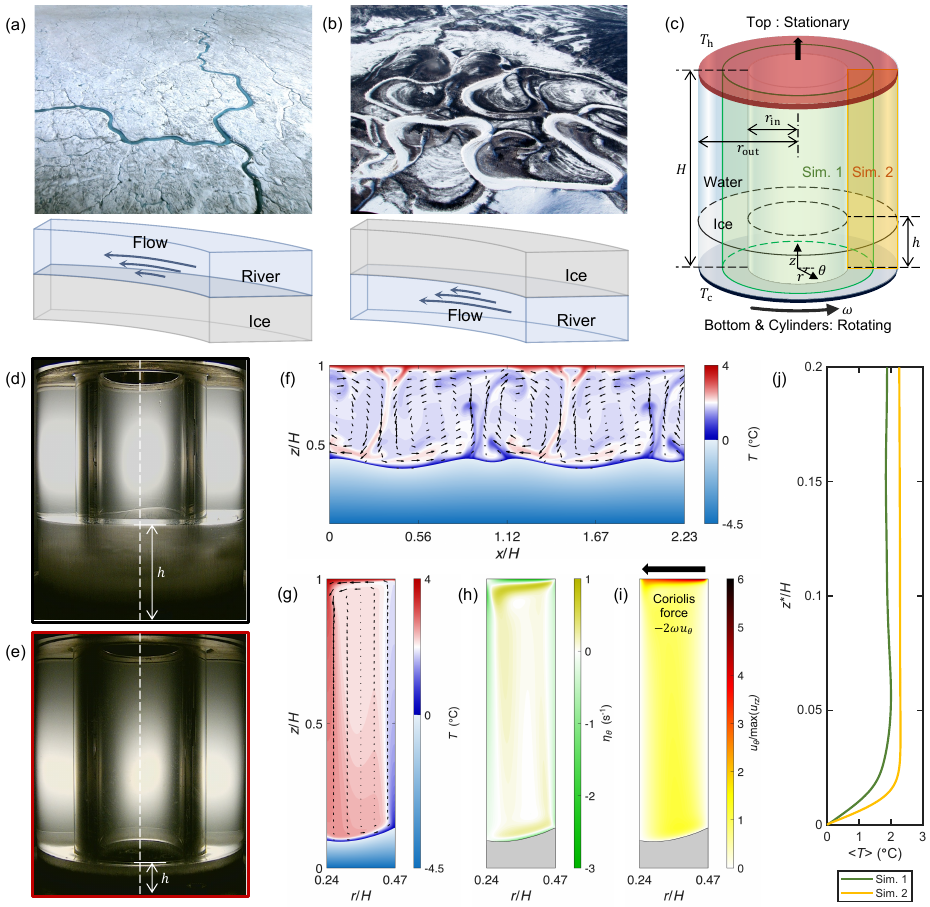}
\caption{Coupled effect of shear and rotation on ice-water phase change.
(a) Meandering meltwater river on the Greenland Ice Sheet, photo by M.-J. Vi{\~n}as/NASA \cite{vinas2015greenland}. 
(b) Frozen meandering river in Alaska, photo by A. Robertson/USFWS (CC BY 2.0) \cite{Robertson2013FrozenRiver}.
The sketches below the photos in (a,b) illustrate shear by the ice-water interface and rotation due to river curvature.
(c) Sketch for the experiment and simulation setup. 
Experiments adopt an annular region with height $H$, inner and outer radii $r_{\textrm{in}}$ and $r_{\textrm{out}}$.
The top hot boundary is stationary and at $T_{\textrm{h}}=4\ ^\circ$C.
The bottom cold rotating boundary is at $T_{\textrm{c}}$, where ice forms.
Inner and outer cylinders rotate in sync with the bottom plate at angular velocity $\omega$.
At $\textrm{Re}=0$, Sim.~1 represents the $\theta$-$z$ cross-section at the mean radius (green surface), omitting curvature and radial variations.
Half of the cross-section is simulated to reduce computational costs.
At $\textrm{Re}=1590.9$, Sim.~2 represents the $r$-$z$ cross-section (yellow surface), with curvature effect and $u_\theta$ retained.
(d,e) Images of equilibrium ice layer with $T_{\textrm{c}}=-4.5\ ^\circ$C and $\textrm{Re}=\omega(r_{\textrm{in}}+r_{\textrm{out}})H/(2\nu)=$0 (d) and 1590.9 (e), where $\nu$ is water kinematic viscosity.
The border colors correspond to the data points in the figures below.
(f) Temperature $T$ (color) and velocity (vector) fields in Sim.~1. 
The horizontal axis is labeled $x$, corresponding to $0.5(r_{\textrm{in}}+r_{\textrm{out}})\theta$, instead of $\theta$ because we neglect the curvature in this simulation.
The result is duplicated in $x$ to represent the full cross-section.
(g-i) Temperature $T$ (color in g), velocity in the $r$-$z$ plane $u_{rz}$ (vector in g), azimuthal vorticity $\eta_\theta=(\partial u_z)/(\partial r)-(\partial u_r)/(\partial z)$ (color in h), and azimuthal velocity $u_\theta$ normalized by the maximum $u_{rz}$ (color in i) in Sim.~2.
(j) Averaged value of temperature at the same distance $z^*$ away from the ice-water interface $\langle T\rangle$, weighted by $r$ in Sim.~2 to account for curvature.
}
\label{fig1}
\end{figure*}

The fluid mechanics of ice-water phase change is attracting significant attention, for its wide relevance in Cryospheric processes \cite{du2024physics}.
Rivers provide a remarkable example of the interaction between ice and flowing water. 
On the ice sheet and ice shelf, meltwater forms surface rivers \cite{pitcher2019supraglacial} (Fig.~\ref{fig1}a), affecting the ice surface mass balance and stability in response to climate change \cite{smith2017direct, tedstone2022increasing, bell2017antarctic}.
Seasonal formation of river ice cover in cold regions changes the means of transportation in winter \cite{brown2018changing} and its melting in spring brings flooding risks \cite{yu2025ice}.
The evolution of an ice-water interface is determined by the fluid dynamics in the neighboring water.
Buoyancy is one of the most frequently studied factors governing the interactions between ice and water \cite{wang2021growth, couston2021turbulent, yang2025asymmetric}.
When other factors are coupled with buoyancy, the complexity of the physical processes will largely increase, bringing diverse phase change phenomena.

In the presence of a mean flow (rivers, ocean currents, etc.), shear is expected to modify the flow structure near the ice-water interface and change the local heat transfer (see the sketches in Fig.~\ref{fig1}a,b).
Different levels of competition and coupling between shear and buoyancy can affect the interface stability, roughness and morphology \cite{toppaladoddi2019combined, guo2025nonlinear}.
For example, a throughflow jet and the associated circulation were found to induce scallops, runnels and ripples on the basal crevasse of the Ross Ice Shelf \cite{washam2023direct}. 
Recently, three-dimensional numerical simulations resolved the formation of ice-water interface ripples overlapping with streamwise patterns over turbulent water streams, advancing the understanding of the multi-scale ice morphodynamics in turbulence \cite{perissutti2024morphodynamics, perissutti2025time}.

For freezing and melting in rivers, the rotation effect can be important because rivers usually meander, for example see Fig.~\ref{fig1}(a,b).
Similar examples in addition to rivers include Earth's rotation \cite{sergienko2013basal} and rotating boundary layer flow in the ocean \cite{waahlin2024swirls}.
Numerical simulation shows that rotation results in columnar vortices that efficiently transport heat and sculpt their structure on the ice layer \cite{ravichandran2021melting}.
Further, rotation and shear are often coupled, together with buoyancy, the flow structure and ice-water interface morphology experience rich transitions depending on their relative intensities \cite{ravichandran2022combined}. Acoustic Doppler current profiler measurements observed helical flow cells under ice around the apex of ice-covered river bends \cite{lotsari2017effects}.

Much of the recent progress in understanding the effect of shear or rotation on ice-water phase change dynamics has come from numerical simulations, or focused on the transient melting behavior with sloping or vertical ice layers (for example \cite{gayen2016simulation, mondal2019ablation, bushuk2019ice}).
In this work, we propose and develop an annular system and study the combined effect of shear and rotation on the equilibrium state of ice-water phase change.
We reveal that shear and rotation couple to induce a secondary circulation perpendicular to the mean flow.
This secondary flow significantly influences the global and local heat transfer, and thereby affects the mean position and morphology of the ice-water interface.
We propose an integral thermal balance model and a local thermal balance model that explain these phenomena well with proper parameterizations.
The local model is extended to real river ice cover and exhibits good qualitative interpretive ability.
This work provides new evidence and understanding for ice-water phase change in the presence of shear and rotation, with relevance to freezing and melting in natural rivers and other curved flows.

\section{II. Experiment and simulation setup} 
We investigate the coupled effect of shear and rotation on the equilibrium ice thickness and ice-water interface morphology in an annular system (Fig.~\ref{fig1}c).
The effective experimental region has a height of $H=0.19$ m, inner and outer radii of $r_{\textrm{in}}=0.045$ m and $r_{\textrm{out}}=0.09$ m. 
Ice grows from the bottom plate, which is set at temperature $T_{\textrm{c}}<0\ ^\circ$C.
The bottom plate rotates with constant angular velocity $-\omega$ (minus sign denotes anti-clockwise direction) together with the inner and outer cylinders.
The top plate is kept at the maximum-density temperature of water $T_{\textrm{h}}=4\ ^\circ$C.
It is held stationary, providing buoyancy and shear to the system.
The corresponding total Rayleigh number $\textrm{Ra}=(\rho_{\textrm{h}}-\rho_0)gH^3/(\rho_0\nu\kappa)$ is $3.82\times10^7$, where $\rho_{\textrm{h}}$ is the water density at $T_{\textrm{h}}$, $\rho_{\textrm{0}}$ is the water density at the freezing/melting point $T_{\textrm{0}}$, $g$ is gravitational acceleration, $\nu$ is water kinematic viscosity and $\kappa$ is water thermal diffusivity.
Experiments are conducted at $T_{\textrm{c}}=-1.6$ and $-4.5\ ^\circ$C at five angular velocities, $\omega=$0, 0.1, 0.5, 1 and 2 rpm.
The corresponding total wall Reynolds number $\textrm{Re}=\omega(r_{\textrm{in}}+r_{\textrm{out}})H/(2\nu)$ spans from 0 to 1590.9. 
See supplementary materials for more information about experimental setup, measurement methods and controlling parameters.

We also perform two numerical simulations with $T_{\textrm{c}}=-4.5\ ^\circ$C to help understand the detailed flow structures and transport situations.
At $\textrm{Re}=0$, Sim.~1 represents the azimuthal-vertical ($\theta$-$z$) cross-section at the mean radius (green surface in Fig.~\ref{fig1}c), omitting curvature and radial variations.
Owing to computational time and cost constraints, it focuses on a rectangular region of length $\pi(r_{\textrm{in}}+r_{\textrm{out}})/2$ (half the cross-section) and height $H$, with stationary top and bottom boundaries and periodic vertical boundaries.
At $\textrm{Re}=1590.9$, Sim.~2 assumes azimuthal symmetry and represents the radial-vertical ($r$-$z$) cross-section (yellow surface in Fig.~\ref{fig1}c).
Although all physical properties depend only on $r$ and $z$, the curvature effect and azimuthal velocity ($u_\theta$) are retained in Sim.~2.
We adapt the code from the \textsc{ch4-project} code.
More details on simulations are provided in the supplementary materials.

\section{III. Ice thinning and secondary circulation by shear and rotation} 
Considering the effects of shear and rotation on the freezing phenomena, one of the most important questions to ask is how they affect the equilibrium ice thickness.
The ice-water interface position is determined by the heat transfer in ice and water \cite{wang2021growth, du2023sea}.
Previous studies show that shear should suppress heat transfer of buoyancy-driven convection \cite{scagliarini2014heat, blass2020flow} in the investigated parameter regime.
The heat transfer response to rotation depends on the flow regime, however, the maximum possible enhancement is still rather limited \cite{ecke2023turbulent}.
One may therefore expect that the ice layer with stronger shear and rotation would be thicker, or at most slightly thinner, simply superposing the separate effects of shear and rotation.
However, as shown in Fig.~\ref{fig1}(d,e), our experiments reveal a substantial reduction in ice thickness against this intuition.
When $\omega=2$ rpm ($\textrm{Re}=1590.9$), the ice layer is only about one-third to one-fifth as thick as when $\omega=0$ rpm ($\textrm{Re}=0$) for both $T_{\textrm{c}}$. 
Since heat is conducted in ice, this large reduction implies a substantially enhanced heat transfer in water.
What is the physical mechanism of this enhancement?

Comparing the results of Sim.~1 and Sim.~2 could provide a mechanism to interpret how shear and rotation modify the flow structure and change the phase change behavior.
Figure~\ref{fig1}(f) shows the temperature (color) and velocity (vector) fields obtained in Sim.~1 ($\textrm{Re}=0$), duplicated in $x$ to represent the full cross-section.
The horizontal axis is labeled $x$, which should correspond to $0.5(r_{\textrm{in}}+r_{\textrm{out}})\theta$ instead of $\theta$ because we neglect the curvature in the simulation.
Buoyancy-driven convection forms convection rolls in the $\theta$-$z$ cross-section.
This implies that curvature and variations in radius ($r$) direction are more or less negligible, at least in determining the ice thickness.
By contrast, in Sim.~2 ($\textrm{Re}=1590.9$), there forms a strong anti-clockwise circulation that spans across the water layer in the $r$-$z$ cross-section, as shown by the vectors in Fig.~\ref{fig1}(g).
Warm water descends along the inner cylinder and cold water rises along the outer cylinder.
Figure~\ref{fig1}(h) reports the azimuthal vorticity, $\eta_\theta=(\partial u_z)/(\partial r)-(\partial u_r)/(\partial z)$.
Adhering to the top plate, there exists a thin shearing layer with very large vorticity opposed to that of the large-scale circulation in the $r$-$z$ cross-section.
This provides a strong pumping effect that drives the circulation.

The formation of this phenomenon can be explained by the B\"{o}dewadt flow \cite{launder2010laminar}.
In the rotating reference frame of angular velocity $-\omega$, the water layer is sheared by the top boundary and the relative azimuthal velocity $u_\theta (r)$ increases vertically from 0 to $\omega r$ at the top boundary, see Fig.~\ref{fig1}(i).
Therefore, the Coriolis force $-2\omega u_\theta$ that points in $-r$ direction overweighs the curvature resistance $u_\theta^2/r$.
Their difference also increases in $+z$ direction and reaches its maximum in the top boundary layer (see Fig.~\ref{fig1}i).
Water is then driven to flow from the outer cylinder to the inner cylinder in the top boundary layer, and further forms the anti-clockwise circulation due to continuity. 
This secondary circulation carries warmer water towards the inner cylinder and colder water towards the outer cylinder.
The formation mechanism of this secondary circulation can also be viewed equivalently from the perspective of the laboratory stationary reference frame.
In the laboratory frame, the magnitude of azimuthal velocity $|u_\theta (r)-\omega r|$ decreases towards the stationary top boundary.
The radial pressure gradient, which is set by the rotation in the bulk, exceeds the local centrifugal support.
Over the temperature range here, the density anomaly of water means that warmer water is denser and colder water is lighter.
The density variation favors the downward motion of warmer water and the upward motion of colder water under earth gravity, in turn strengthening this circulation.
Compared to Sim.~1, the mean temperature in water and the temperature gradient in the bottom thermal boundary layer are both higher in Sim.~2 (Fig.~\ref{fig1}j).
These features suggest that the secondary circulation in $r$-$z$ cross-section enhances heat transfer in water and explains the thinner equilibrium ice layer.

\begin{figure}[!tb]
\centering
\includegraphics[width=1\linewidth]{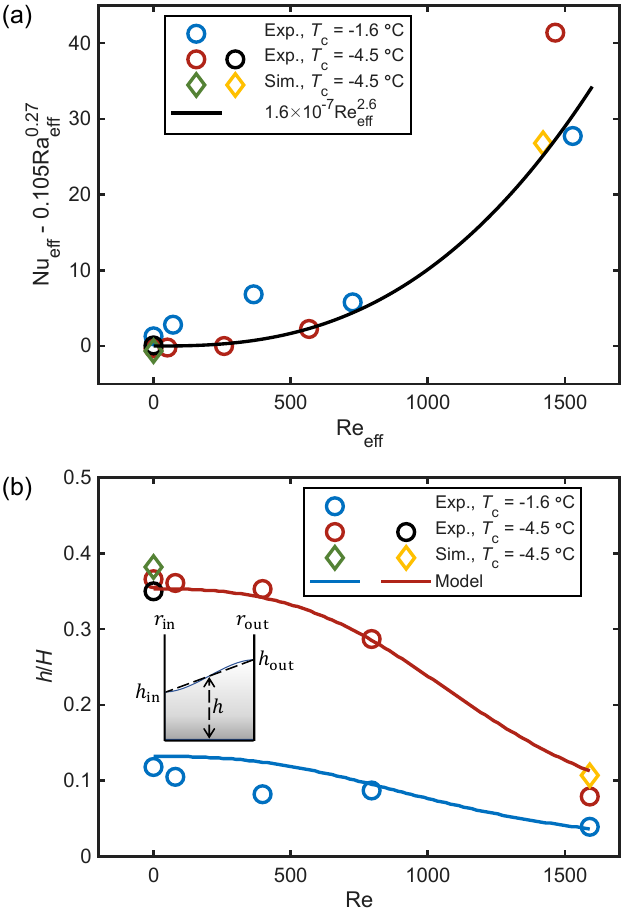}
\caption{Integral thermal balance model for heat transfer and mean ice thickness.
The blue circles are for experiments with $T_{\textrm{c}}=-1.6\ ^\circ$C, the red circles and black circle are for experiments with $T_{\textrm{c}}=-4.5\ ^\circ$C, and the green and yellow diamonds are for Sim.~1 and Sim.~2 with $T_{\textrm{c}}=-4.5\ ^\circ$C.
(a) Inferred additional heat transfer efficiency $\textrm{Nu}_{\textrm{eff}}-0.105\textrm{Ra}_{\textrm{eff}}^{0.27}$ as a function of effective wall Reynolds number $\textrm{Re}_{\textrm{eff}}$. $\textrm{Nu}_{\textrm{eff}}$ is the effective Nusselt number calculated with global thermal balance with experiment and simulation results.
$\textrm{Ra}_{\textrm{eff}}$ is the effective Rayleigh number. 
The black line shows the result estimating the heat transfer enhancement with a power-law of $\textrm{Re}_{\textrm{eff}}$.
(b) Model results (blue line for $T_{\textrm{c}}=-1.6\ ^\circ$C and red line for $T_{\textrm{c}}=-4.5\ ^\circ$C) on the mean ice thickness $h$ compared with experiment and simulation results.
In the model, $\textrm{Nu}_{\textrm{eff}}$ is estimated with $0.105\textrm{Ra}_{\textrm{eff}}^{0.27}+1.6\times 10^{-7}\textrm{Re}_{\textrm{eff}}^{2.6}$.
The inset shows that $h$ in the experiments and simulations is approximated as the average of ice thicknesses on the inner and outer cylinder.
}
\label{fig2}
\end{figure}

\section{IV. Integral thermal balance model} 
Next, we propose a model for the equilibrium ice thickness by examining the global thermal balance at the ice-water interface and connect the thinning of ice to the additional heat transfer due to shear and rotation.
The key is to scale the effective heat transfer efficiency with the effective controlling parameters.

With conductive heat transfer through the ice, the global thermal balance at the ice-water interface writes:
\begin{equation}
q=-\Lambda_{\textrm{ice}}\frac{T_0-T_{\textrm{c}}}{h}=-\textrm{Nu}_{\textrm{eff}}\Lambda_{\textrm{water}}\frac{T_{\textrm{h}}-T_0}{H-h},
\label{global}
\end{equation}
where $q$ is the mean heat flux across the system, $\Lambda_{\textrm{ice}}$ and $\Lambda_{\textrm{water}}$ are thermal conductivities of ice and water, $T_0$ is freezing/melting point, and the effective Nusselt number $\textrm{Nu}_{\textrm{eff}}$ estimates the dimensionless heat transfer efficiency in water.
As opposed to the total parameters, the effective controlling parameters use the remaining depth of water as the typical length scale.
They are the effective buoyancy intensity $\textrm{Ra}_{\textrm{eff}}=(\rho_{\textrm{h}}-\rho_0)g(H-h)^3/(\rho_0\nu\kappa)$ and the effective shear/rotation intensity $\textrm{Re}_{\textrm{eff}}=\omega(r_{\textrm{in}}+r_{\textrm{out}})(H-h)/(2\nu)$.

Based on the discussion in the previous section, we describe  $\textrm{Nu}_{\textrm{eff}}$ by a buoyancy-based background contribution $\textrm{Nu}_{\textrm{buo}}$ and an additional contribution $\textrm{Nu}_{\textrm{add}}$.
$\textrm{Nu}_{\textrm{buo}}$ depends on $\textrm{Ra}_{\textrm{eff}}$, reflecting the background heat transfer due to buoyancy-driven convection in the azimuthal direction.
$\textrm{Nu}_{\textrm{add}}$ depends on $\textrm{Re}_{\textrm{eff}}$, accounting for the overall additional heat transfer resulting from shear, rotation and related secondary effects.
We estimate that:
\begin{equation}
\begin{aligned}
 \textrm{Nu}_{\textrm{eff}}&=\textrm{Nu}_{\textrm{buo}}(\textrm{Ra}_{\textrm{eff}})+\textrm{Nu}_{\textrm{add}}(\textrm{Re}_{\textrm{eff}})\\    &=0.105\textrm{Ra}_{\textrm{eff}}^{0.27}+1.6\times10^{-7}\textrm{Re}_{\textrm{eff}}^{2.6},
    \label{Nueff}
\end{aligned}
\end{equation}
where the exponent 0.27 is adopted from previous studies \cite{wang2021growth, du2023sea} on icing with buoyancy-driven convection in rectangular systems.
The other coefficients and the exponent 2.6 are fitted with our experiment and simulation results.
Finally, combining Eqs.~\ref{global} and \ref{Nueff}, we are able to obtain the equilibrium ice thickness $h$ under different temperature and rotation conditions.

Since it is difficult to reconstruct the three-dimensional ice-water interface in the experiments, the mean ice thickness $h$ is estimated approximately with the average of ice thicknesses on the inner and outer cylinders.
Figure~\ref{fig2}(a) compares experiment and simulation results on the inferred additional heat transport (circles and diamonds), $\textrm{Nu}_{\textrm{eff}}-0.105\textrm{Ra}_{\textrm{eff}}^{0.27}$ where $\textrm{Nu}_{\textrm{eff}}$ are obtained with Eq.~\ref{global} from experiments and simulation, with the empirical $\textrm{Nu}_{\textrm{add}}$ (the second term in Eq.~\ref{Nueff}, black line).
It can be seen that the model captures the overall trend of the additional heat transfer efficiency as a function of $\textrm{Re}_{\textrm{eff}}$.
Figure~\ref{fig2}(b) shows the experiment and simulation results of $h$ (circles and diamonds), and compares them with the model results calculated by substituting Eq.~\ref{Nueff} into Eq.~\ref{global} (lines).
As expected, $h$ decreases significantly with increasing shear and rotation.
The simulated $h$ are broadly consistent with the corresponding experimental results (Fig.~\ref{fig2}b).
This agreement suggests that the two reduced simulations capture the flow structures relevant to the global heat transfer and mean ice thickness, although they do not reconstruct the full three-dimensional flow field.

These consistencies provide evidence for our previous discussions on the physical mechanism and our modelling: The combined effect of shear and rotation leads to circulation in $r$-$z$ cross-section, enhancing heat transfer in addition to the background convection.

\section{V. Ice morphology and local thermal balance model} 

\begin{figure*}[!htb]
\centering
\includegraphics[width=0.85\linewidth]{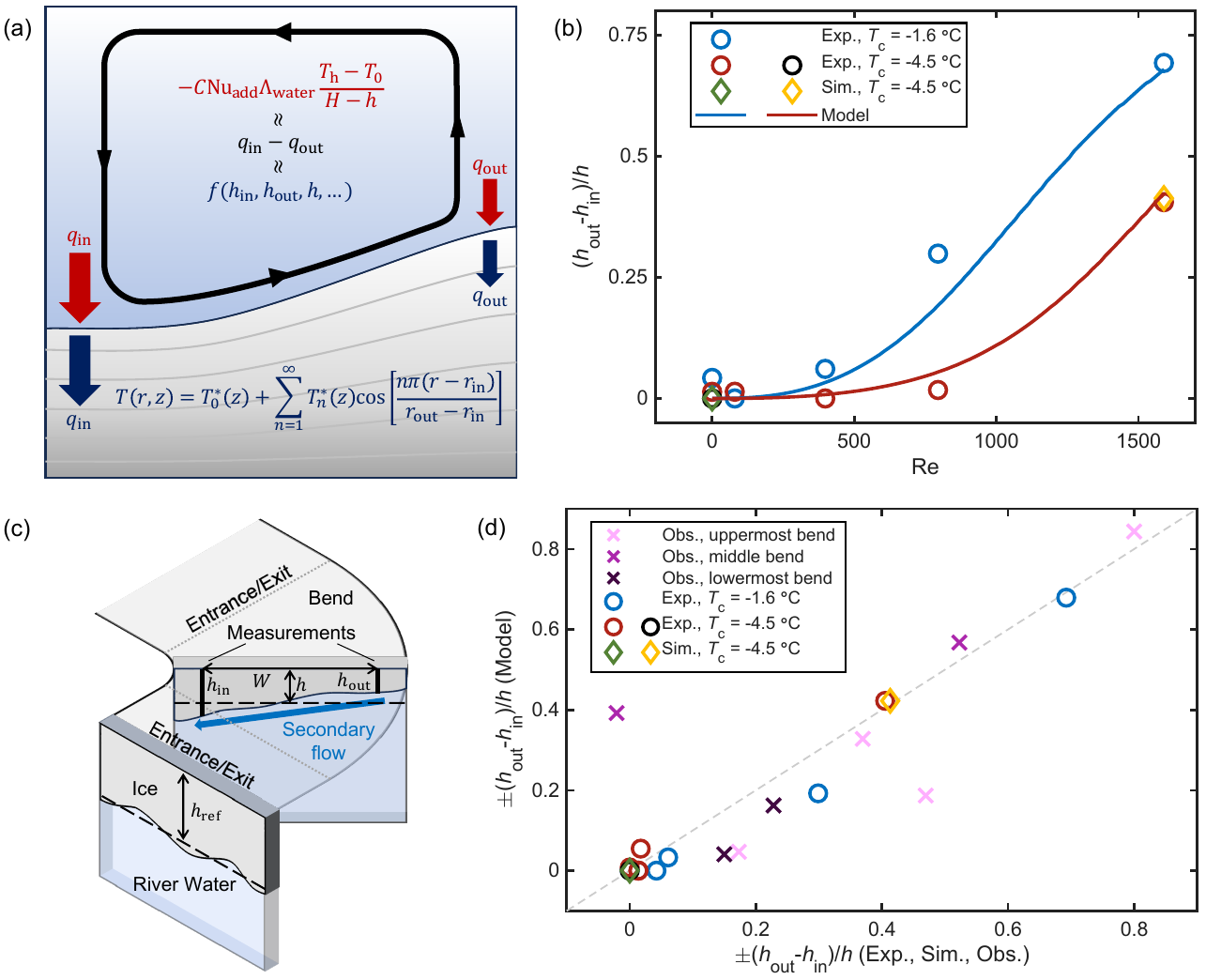}
\caption{Local thermal balance model for the inner-outer ice thickness difference and its application to explain the transversal ice thickness difference in the meandering river bend.
In b and d, the blue circles are for experiments with $T_{\textrm{c}}=-1.6\ ^\circ$C, the red circles and black circle are for experiments with $T_{\textrm{c}}=-4.5\ ^\circ$C, and the green and yellow diamonds are for Sim.~1 and Sim.~2 with $T_{\textrm{c}}=-4.5\ ^\circ$C.
(a) Conceptual schematic for the model. The model obtains the inner-outer ice thickness difference by relating it to the heat flux difference which is estimated to be proportional to the enhancement of heat transfer due to rotation and shear, with modal decomposition of temperature in ice.
(b) Model results (blue line for $T_{\textrm{c}}=-1.6\ ^\circ$C and red line for $T_{\textrm{c}}=-4.5\ ^\circ$C) on the inner-outer ice thickness difference normalized by the mean ice thickness $(h_{\textrm{out}}-h_{\textrm{in}})/h$ compared with experiment and simulation results. 
(c) Conceptual schematic of the water flow and ice cover near river bend.
Ice cover provides shear to the rotating flow in river bend, which potentially gives rise to a secondary flow towards the inner bend (blue arrow) and leads to the transversal ice thickness difference. 
The entrance or exit whose river depth is comparable to that of the bend is selected as reference for estimating the additional heat transfer at the bend.
(d) Comparison on normalized transversal/radial ice thickness difference between results from the model and from field observations, experiments and simulations.
The purple crosses show the comparison at three bends of Pulmanki River, with the thickness difference defined as $-(h_{\textrm{out}}-h_{\textrm{in}})/h$.
The circles and diamonds show the comparison in our annular system, with the thickness difference defined as $(h_{\textrm{out}}-h_{\textrm{in}})/h$.
Data of the Pulmanki River are obtained from Ref.~\cite{kamari2017spatial}.
}
\label{fig3}
\end{figure*}

Aside from the mean thickness, an even more interesting result is the ice-water interface morphology: The rising of the ice-water interface in positive $r$ direction becomes more pronounced as shear and rotation intensify.
It is difficult to reconstruct the ice-water interface in the experiments, but difference in ice thicknesses at the inner and outer cylinders can reflect such variation.
With zero or small $\textrm{Re}$, the inner and outer ice layers are almost of the same thickness (see Fig.~\ref{fig1}d).
With high $\textrm{Re}$, however, the outer ice layer is obviously thicker than the inner ice layer (see Fig.~\ref{fig1}e).
Sim.~2 also finds that local ice thickness generally increases in positive $r$ direction (see Fig.~\ref{fig1}g-i).

It is understandable that this ice-water interface morphology would form.
As discussed above, the coupled effect of shear and rotation promotes the formation of anti-clockwise large-scale circulation in the $r$-$z$ cross-section.
Water flows from the inner cylinder to the outer cylinder in the boundary layer adhering to the ice layer.
Similar to the cases in Refs.~\cite{wang2021ice, wang2021equilibrium, yang2023morphology}, such a development of boundary layer leads to difference in heat flux, thereby shaping the ice-water interface.
Here, we propose a model for the difference between inner ice thickness $h_{\textrm{in}}$ and outer thickness $h_{\textrm{out}}$ by estimating the difference between local heat flux at inner ice-water interface $q_{\textrm{in}}$ and outer interface $q_{\textrm{out}}$, as illustrated in Fig.~\ref{fig3}(a).

First, we relate $h_{\textrm{out}}-h_{\textrm{in}}$ with $q_{\textrm{in}}-q_{\textrm{out}}$ through modal decomposition of temperature in ice.
As the ice thickness varies in the radial direction, local temperature also varies between the inner and outer cylinders, with nearly adiabatic boundaries.
Therefore, local temperature in ice can be approximated with cosine modes in the radial direction, as done in Ref.~\cite{couston2021topography}:
\begin{equation} T(r,z)=T_0^*(z)+\sum_{n=1}^\infty T_n^*(z)\textrm{cos}\left[\frac{n\xi(r-r_{\textrm{in}})}{H}\right],
\label{Tdecompose}
\end{equation}
where $\xi=\pi H/(r_{\textrm{out}}-r_{\textrm{in}})$.

To compare and then estimate temperatures at equal distances below the inner and outer ice-water interface, we define $\Delta(z)=T(r_{\mathrm{in}},z)-T(r_{\mathrm{out}},z+h_{\textrm{out}}-h_{\textrm{in}})$. 
We consider the zero-order and first-order components of $T$ and approximate the zero-order component with the mean temperature gradient in ice.
The interface condition gives $\Delta(h_{\textrm{in}})=0$.
Another boundary condition is $\Delta(0)=-(T_0-T_{\textrm{c}})(h_{\textrm{out}}-h_{\textrm{in}})/h$, by approximating the temperature gradient near the cold boundary with the mean value.
These analyses, together with local energy conservation at the inner and outer interface, give $\Delta$:
\begin{equation}
    \Delta=\frac{h_{\textrm{out}}-h_{\textrm{in}}}{h}(T_0-T_{\textrm{c}})\left[\frac{\textrm{sinh}(\xi z/H)}{\textrm{sinh}(\xi h_{\textrm{in}}/H)}-1\right].
    \label{Delta}
\end{equation}

Next, we can get the corresponding difference in conductive heat flux at the inner and outer ice-water interface on the ice side with Eq.~\ref{Delta}.
On the water side, we assume that the background heat transfer due to buoyancy-driven convection in the
azimuthal direction is uniform in the radial direction.
The additional heat transfer due to circulation in $r$-$z$ cross-section has a certain distribution in the radial direction.
Under these two assumptions, the difference in convective heat flux at the inner and outer ice-water interface on the water side should scale with the additional heat transfer associated with the secondary circulation $\textrm{Nu}_{\textrm{add}}$.

Finally, approximating the $h_{\textrm{in}}$ in the denominator of Eq.~\ref{Delta} with $h$ and combining Eq.~\ref{global}, we get:
\begin{equation}
    \frac{h_{\textrm{out}}-h_{\textrm{in}}}{h}=C\frac{\textrm{Nu}_{\textrm{add}}H}{\textrm{Nu}_{\textrm{eff}}\xi h}\textrm{tanh}\left(\frac{\xi h}{H}\right),
    \label{local}
\end{equation}
where $C>0$ is an empirical coefficient relating the inner-outer heat flux difference to the additional heat transfer, and $h$, $\textrm{Nu}_{\textrm{add}}$ and $\textrm{Nu}_{\textrm{eff}}$ can be estimated with Eqs.~\ref{global} and \ref{Nueff}.
Detailed derivation of the model is provided in the Appendix. 

Figure \ref{fig3}(b) compares the experiment and simulation results (circles and diamonds) on $(h_{\textrm{out}}-h_{\textrm{in}})/h$ with the model results (lines), choosing $C=1$.
It can be seen that the model obtains well the thickness difference between inner and outer ice layer at various temperature, shear and rotation conditions.

Overall, experiments, simulations and the local thermal balance model provide a consistent understanding of the ice-water interface morphology: The direction and intensity of the circulation induced by shear and rotation determine the non-uniformity of the local heat flux in $r$ direction, leading to the variation in local ice thickness.

Equations~\ref{global} and \ref{local} describe two different consequences resulting from the same fluid dynamical mechanism.
Secondary circulation due to coupled shear and rotation brings non-uniform heat transfer enhancement in the radial direction.
Enhanced mean heat transfer reduces the mean equilibrium ice thickness, while radially unequal enhancement produces non-uniform ice thickness and thereby shapes the ice-water interface.

\section{VI. Application to geophysical phenomena}

In this section, we take the ice cover in a subarctic meandering river as an example to illustrate how the discussed physical mechanism and proposed model help explain geophysical phenomena in the Cryosphere.
The river of focus here is a roughly 1.2 km part of the Pulmanki River (69$^\circ$ 55$^\prime$ 50$^{\prime\prime}$ N, 28$^\circ$ 2$^\prime$ 24$^{\prime\prime}$ E) in Northern Finland.
Ref.~\cite{kamari2017spatial} measures the spatial ice thickness variation in three meandering bends of this river in February, 2014.
Excluding the part of ice cover that is completely frozen and restricted by terrain, ice is thicker on the inner side of the bend than that on the outer side in most of the cross-sections, which is analogous to the model system discussed above.

The bend curves the streamlines and the ice cover above the river provides shear.
Therefore, curvature-induced Coriolis force (distinct from Earth-rotation Coriolis force) can drive secondary water flow from the outer bend to the inner bend adjacent to the ice cover, sculpturing an ice-water interface that is lower on the inside and higher on the outside, as shown in Fig.~\ref{fig3}(c).
Explaining the transversal ice thickness difference with Eq.~\ref{local} requires estimating by how much shear and rotation enhance heat transfer.
It is difficult to accurately determine this ratio based on measurements or fluid-dynamical calculations.
To adapt our model, we assume that the background heat flux in water and the heat transfer in ice are similar among the cross-sections in the bends and at the entrance/exit with similar river depth, and therefore $\textrm{Nu}_{\textrm{add}}/\textrm{Nu}_{\textrm{eff}}\sim1-h/h_{\textrm{ref}}$ where $h$ is the mean ice thickness in the bend and $h_{\textrm{ref}}$ is at the reference entrance/exit (see Fig.~\ref{fig3}c).
We incorporate the coefficient of this proportionality into $C$ and estimate the transversal ice thickness difference with:
\begin{equation}
    \frac{h_{\textrm{in}}-h_{\textrm{out}}}{h}=C\left(1-\frac{h}{h_{\textrm{ref}}}\right)\frac{W}{\pi h}\textrm{tanh}\left(\frac{\pi h}{W}\right).
    \label{geo}
\end{equation}

We compare the observation and model results on the transversal ice thickness difference in the bends with the borehole measurement data reported in Ref.~\cite{kamari2017spatial}.
As shown in Fig.~\ref{fig3}(c), for the observation results, $h_{\textrm{in}}$ and $h_{\textrm{out}}$ are the innermost and outermost borehole thicknesses, and $h$ is the mean ice thickness.
For the model results, $h_{\textrm{ref}}$ is the mean ice thickness at the bend entrance or exit whose river depth is more similar to that of the bend, $W$ is the distance between the two boreholes, and $C$ is set as 2.5 by fitting the observation results.

The results, four from the uppermost bend, two from the middle bend and two from the lowermost bend, are presented as the purple crosses in Fig.~\ref{fig3}(d), see more on the location of the adopted cross-sections and reference cross-sections in the supplementary materials.
Our laboratory results are also attached.
The data points are roughly aligned with a slope of 1, despite the presence of an outlier, which implies that our model can qualitatively explain the observed transversal ice thickness difference, and that the role of secondary circulation deserves further quantification.

\section{VII. Conclusion and outlook}

Combining experiments, numerical simulations and theoretical modelling, we investigate the combined effects of shear and rotation on the coupled dynamics of ice-water phase change and convective flow in an annular system.
Shear and rotation lead to a maximum inward forcing (difference between Coriolis force and curvature resistance) near the moving boundary in the rotating reference frame (stationary in the laboratory frame).
It induces a Bödewadt-type pumping, driving the formation of a secondary circulation in the cross-section perpendicular to the main flow direction.
The resulting enhancement of global heat transfer substantially reduces the mean equilibrium ice thickness, while non-uniform interfacial heat flux determines the variation in local ice thickness in the radial/transversal direction and thereby the ice-water interface morphology.

An integral thermal balance model and a local thermal balance model are proposed to explain the system responses to shear and rotation.
The integral model considers the mean effective heat transfer efficiency in water as the sum of a background buoyancy contribution and an additional contribution due to shear and rotation.
The mean equilibrium ice thickness is then determined with the balance between global conductive heat transfer in ice and convective heat transfer in water.
The local model relates the inner-outer ice thickness difference to the spatially non-uniform additional heat transfer in water, via modal decomposition of temperature in ice and local heat flux balance at the ice-water interface.
The models capture well the experiment and simulation results by appropriately parameterizing the heat transfer with effective controlling parameters.
The local model can further be adapted to ice-covered meandering rivers, yielding qualitative agreement with published measurements of transverse ice thickness variations.

Together, these results reveal a common physical mechanism on how shear and rotation change the flow structure, affect heat transfer magnitude and spatial distribution, and influence the ice-water interface position and morphology.
Despite adopting an idealized system, the connection with river-ice observations indicates the geophysical relevance of our findings.
This motivates detailed field observation in rivers and other related geophysical situations to further quantify the contribution of secondary circulation induced by shear and rotation to ice evolution.
Further work is encouraged to link the flow structure, heat transfer and phase change dynamics across a broader range of conditions, such as when shear and rotation are coupled with salinity in double-diffusive convection \cite{wilson2023double, guo2025effects, rosevear2021role}, with radiation \cite{yang2023bistability}, with vibration \cite{fang2024vibration}, and with different inclusions in ice \cite{wengrove2023melting, du2025sea, wang2026freezing}.
Different system geometries also deserve to be tested, especially when the ice-water interface is tilted or vertical \cite{yang2024enhanced, weady2022anomalous}.
Simultaneous experimental measurements and three-dimensional simulations would help develop this description in a more robust and universal manner. 

\section{Acknowledgements} 
This work is supported by NSFC Excellence Research Group Program for `Multiscale Problems in Nonlinear Mechanics' (No. 12588201), the Natural Science Foundation of China under Grant Nos. 12602337 and 12402321, the New Cornerstone Science Foundation through the New Cornerstone Investigator Program and the XPLORER PRIZE, the China Postdoctoral Science Foundation No. 2026M791982, the Postdoctoral Fellowship Program of CPSF No. GZC20261056, and Shuimu Tsinghua Scholar Program under Grant No. 2023SM038.

\section{Appendix: Detailed derivation of the local model}
\setcounter{equation}{0}
\renewcommand{\theequation}{A\arabic{equation}}

As mentioned above, we first relate the radial ice thickness difference with the interface heat flux difference.

Assuming symmetry in $\theta$ direction, i.e., $\partial/(\partial\theta)=0$, the local energy conservation in the ice at equilibrium writes:
\begin{equation}
    \frac{\partial^2 T(r,z)}{\partial z^2}+ \frac{\partial^2 T(r,z)}{\partial r^2} +  \frac{\partial T(r,z)}{r\partial r}=0.
    \label{A1}
\end{equation}
Consider the inner and outer cylinders as nearly adiabatic:
\begin{equation}
    \frac{\partial T(r,z)}{\partial r}=0,\ r=r_{\textrm{in}}, r_{\textrm{out}}.
    \label{A2}
\end{equation}
Eq.~\ref{A2} indicates that we can do the modal decomposition of $T(r,z)$ as Eq.~\ref{Tdecompose}:
\begin{equation} T(r,z)=T_0^*(z)+\sum_{n=1}^\infty T_n^*(z)\textrm{cos}\left[\frac{n\xi(r-r_{\textrm{in}})}{H}\right],
\label{A3}
\end{equation}
where $\xi=\pi H/(r_{\textrm{out}}-r_{\textrm{in}})$.
Considering the zero-order and first-order components of Eq.~\ref{A3} (Eq.~\ref{Tdecompose}), we have:
\begin{equation}
\begin{aligned}
    \frac{\partial^2 T}{\partial r^2}=&\frac{\partial^2}{\partial r^2}\left[T_1^*\textrm{cos}\frac{\xi(r-r_{\textrm{in}})}{H}\right]\\=&-\left(\frac{\xi}{H}\right)^2\left[T_1^*\textrm{cos}\frac{\xi(r-r_{\textrm{in}})}{H}\right]\\=&-\left(\frac{\xi}{H}\right)^2(T-T_0^*).
    \label{A4}
\end{aligned}
\end{equation}
Eqs.~\ref{A1}, \ref{A2} and \ref{A4} yield:
\begin{equation}
    \frac{\partial^2 T}{\partial z^2}-\left(\frac{\xi}{H}\right)^2(T-T_0^*)=0,\ r=r_{\textrm{in}}, r_{\textrm{out}}.
    \label{A5}
\end{equation}

Defining $\Delta(z)=T(r_{\mathrm{in}},z)-T(r_{\mathrm{out}},z+h_{\textrm{out}}-h_{\textrm{in}})$, we calculate the difference between Eq.~\ref{A5} at $r=r_{\mathrm{in}}$ and $r_{\mathrm{out}}$:
\begin{equation}
      \frac{d^2\Delta}{dz^2}-\left(\frac{\xi}{H}\right)^2\left[\Delta-T_0^*(z)+T_0^*(z+h_{\textrm{out}}-h_{\textrm{in}})\right]=0.
    \label{A6}
\end{equation}
Estimating the zero-order component with the mean temperature gradient in ice, i.e., $T_0^*(z)=T_{\textrm{c}}+(T_0-T_{\textrm{c}})z/h$, we get the Ordinary Differential Equation for $\Delta$:
\begin{equation}
      \frac{d^2\Delta}{dz^2}-\left(\frac{\xi}{H}\right)^2\Delta-\left(\frac{\xi}{H}\right)^2\frac{h_{\textrm{out}}-h_{\textrm{in}}}{h}(T_0-T_{\textrm{c}})=0.  
    \label{A7}
\end{equation}
The interface condition gives $\Delta(h_{\textrm{in}})=0$.
Approximating the temperature gradient near the cold boundary with the mean value, another boundary condition is $\Delta(0)=-(T_0-T_{\textrm{c}})(h_{\textrm{out}}-h_{\textrm{in}})/h$.
Solving Eq.~\ref{A7}, we get Eq.~\ref{Delta}:
\begin{equation}
    \Delta=\frac{h_{\textrm{out}}-h_{\textrm{in}}}{h}(T_0-T_{\textrm{c}})\left[\frac{\textrm{sinh}(\xi z/H)}{\textrm{sinh}(\xi h_{\textrm{in}}/H)}-1\right].
    \label{A8}
\end{equation}

Therefore, the corresponding difference in conductive heat flux at the inner and outer ice-water interface on the ice side $q_{\textrm{in}}-q_{\textrm{out}}$ is:
\begin{equation}
\begin{aligned}
    q_{\textrm{in}}-q_{\textrm{out}}&=-\Lambda_{\textrm{ice}}\left(\frac{\partial T}{\partial z}|_{r_{\textrm{in}},h_{\textrm{in}}}-\frac{\partial T}{\partial z}|_{r_{\textrm{out}},h_{\textrm{out}}}\right)\\&=-\Lambda_{\textrm{ice}}\frac{d\Delta}{dz}|_{z=h_{\textrm{in}}}\\&=-\frac{(h_{\textrm{out}}-h_{\textrm{in}})\Lambda_{\textrm{ice}}\xi(T_0-T_{\textrm{c}})}{hH\textrm{tanh}(\xi h_{\textrm{in}}/H)}.
\end{aligned}
        \label{A9}
\end{equation}

As mentioned above, we assume that the background heat transfer due to buoyancy-driven convection in the
azimuthal direction is uniform in the radial direction.
The additional heat transfer due to circulation in $r$-$z$ cross-section has a certain distribution in the radial direction.
On the water side, we have:
\begin{equation}
    q_{\textrm{in}}-q_{\textrm{out}}\propto -\textrm{Nu}_{\textrm{add}}\Lambda_{\textrm{water}}\frac{T_{\textrm{h}}-T_0}{H-h}.
    \label{A10}
\end{equation}

Finally, approximating the $h_{\textrm{in}}$ in the denominator of Eq.~\ref{A9} with $h$, Eqs.~\ref{A9} and \ref{A10}, together with the global thermal balance (Eq.~\ref{global}), give Eq.~\ref{local}:
\begin{equation}
    \frac{h_{\textrm{out}}-h_{\textrm{in}}}{h}=C\frac{\textrm{Nu}_{\textrm{add}}H}{\textrm{Nu}_{\textrm{eff}}\xi h}\textrm{tanh}\left(\frac{\xi h}{H}\right),
    \label{A11}
\end{equation}
where $C>0$ is an empirical coefficient relating the inner-outer heat flux difference to the additional heat transfer.

\normalem 
\bibliography{ref.bib}

\end{document}